\documentclass{epl}
\title{Spin screening of magnetic moments in superconductors}
\author{F.S.Bergeret\inst{1,4} \and A.F. Volkov\inst{1,2} \and K.B.Efetov\inst{1,3}}
\institute{
\inst{1} Theoretische Physik III,Ruhr-Universit\"{a}t Bochum, D-44780 Bochum, Germany.\\
\inst{2} Institute of Radioengineering and Electronics of the Russian Academy of Sciences, 103907 Moscow, Russia \\
\inst{3} L.D. Landau Institute for Theoretical Physics, 117940
 Moscow, Russia\\
 \inst{4} Laboratorio de F\'{i}sica
de Sistemas Peque${\it \tilde{n}}$os y Nanotecnolog\'{i}a, CSIC,
Serrano144, E-28006 Madrid}


\begin{document}

\maketitle

\begin{abstract}
We consider ferromagnetic particles embedded into a superconductor and study
the screening of their magnetic moments by the spins of the
Cooper pairs in the superconductor. It is shown that a magnetic moment
opposite to the one of the ferromagnetic particle is induced in the
superconductor. In the case of a small itinerant ferromagnet grain and low
temperatures the full screening of the magnetic moment takes place, \textit{%
i.e} the absolute value of the total magnetic moment induced in the
superconductor is equal to the one of the ferromagnetic particle. In type II
superconductors the proposed screening by spins of the conduction electrons
can be much stronger than the conventional screening by Meissner currents.
\end{abstract}

The phenomenon of screening is very common in physics. The best known
example {is the screening of an electric charge in metals due to a
redistribution of free electrons in space. This charge screening is very
strong and the length characterizing an exponential decay of the electric
field (the Thomas-Fermi length $\lambda _{TF}$) is in most metals of the
order of the Fermi wave length, i.e. of the order of the interatomic
spacing. } The length $\lambda _{TF}$ does not depend on whether the metal
is in the normal or in the superconducting state.

Another famous example is the screening of a magnetic field or a magnetic
moment by superconducting currents in a superconductor (Meissner effect){\
\cite{abri_book}}. Due to this effect the magnetic field decays  over
the length  $\lambda _{L}$ (the  penetration depth) and vanishes in a bulk superconductor.
The same length characterizes the decay of the magnetic field created by a
ferromagnetic $(F)$ grain embedded in a superconductor. In contrast to the
screening of the electric charge in normal metals the screening of the
magnetic field and magnetic moments in superconductors is weaker and the penetration depth $%
\lambda _{L}$ can be of the order of hundreds interatomic distances or
larger. {The screening of the magnetic moment is a phenomenon specific for a
superconductor and, in contrast to the charge screening, is very small in a
normal metal. Although the stray field created by a ferromagnetic grain
embedded in a nonmagnetic (}${N}${) metal may induce a negative local
magnetization in some regions of the normal metal due to the Pauli
paramagnetism, the susceptibility is rather small ($\mu _{B}^{2}\nu \sim
10^{-6}$, $\mu _{B}$ is the Bohr magneton and $\nu $ the density of states
at the Fermi level) and the screening can be neglected.}
For certain geometries{\ of a superconducting sample (films, wires) the
penetration length }$\lambda _{L}$ can exceed the transversal size and the
screening of the magnetic moment of a ferromagnetic particle due to the
Meissner currents does not play an essential role.{\ It is usually believed
that in such a situation the total magnetic moment is just the magnetic
moment of the ferromagnetic particle and no additional magnetization is
induced by the electrons of the superconductor. This common wisdom is quite
natural because, at first glance, a possible contribution into the screening
of the electron spins in the superconductor is even smaller than in the
normal metal. In conventional superconductors the total spin of a Cooper
pair is equal to zero and the polarization of the conduction electrons is
even smaller than in the normal metal. Spin-orbit interactions may lead to a
finite magnetic susceptibility of the superconductor \cite{Abr&Gor}, but it
is positive and anyway smaller than in the normal state. }

In this Letter we suggest a new mechanism for the screening of the magnetic
moment of a ferromagnetic particle embedded in a superconductor by spins of
the superconducting electrons. This effect is large and the magnetic moment
of the grain can be completely screened. The characteristic length of the
screening is of the order of the size of the Cooper pair $\xi _{S}=\sqrt{%
D_{S}/2\pi T_{c}}$ (we consider the ``dirty'' limit) and can be much smaller
than the penetration depth $\lambda _{L}$ in type II superconductors. If the
size of the superconductor in the transverse direction is smaller than the
penetration length $\lambda _{L}$, the mechanism we propose is the only one
leading to the screening of the magnetic moment of the ferromagnetic grain.

Although this effect seems very surprising and has been overlooked in all
previous investigations of  superconductivity, its origin can be
understood without any calculations. This additional screening arises due to
the exchange interaction between the spins of the conduction electrons and
the magnetic moment of the $F$ grain, and the possibility for the
superconducting condensate to penetrate the grain.

If the size of the grain is much smaller than the size of the Cooper pair $%
\xi _{S}$ the probability that both  electrons of a Cooper pair are
located in the ferromagnetic grain is small. Therefore one can assume that only
one electron of the Cooper pair spends some time in the grain. Then, the
exchange interaction enforces the spin of this electron to be parallel to
the magnetization in the grain. This leaves no choice for the second
electron in the Cooper pair but to be antiparallel to the magnetization of
the grain. In this way an additional magnetization antiparallel to the one in
the grain is induced in the superconductor. Of course, if the transparency
of the $S/F$ interface is small the induced magnetization is weak. However,
if the transparency is high enough the induced magnetization is large and
the magnetization of the grain can be completely screened.

Below we support this qualitative discussion by an explicit calculation
using quasiclassical Green's functions and a proper Usadel type equation
including the exchange field. Although the calculational scheme is quite
standard, it contains a new important ingredient: the superconducting
condensate in the $F$ region near the surface consists of a singlet and a triplet components \cite{BVE1}. Due to the proximity effect the
triplet component penetrates the superconductor over the length $\xi _{S}$
and, as it carries spin, polarizes the superconductor. The triplet component
was not considered previously for this type of problems.
%%%%%%%%%%%%%%%%%%%%%%%%%%%%%%%%%%
\begin{figure}[tbp]
\onefigure[scale=0.5]{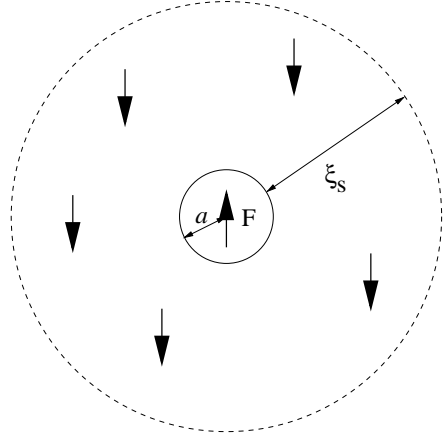}
\caption{Schematic representation of a magnetic grain embedded in a
superconductor. The arrows show the direction of magnetization. In the
superconductor a negative magnetization is induced over distances of the
order of $\protect\xi _{s}$.}
\label{fig.1}
\end{figure}
%%%%%%%%%%%%%%%%%%%%%%%%%%%%%%%%%%%
We consider a ferromagnetic spherical grain of the radius $a$ embedded in a
superconductor $S$ (Fig.\ref{fig.1}). The superconductor is assumed to be a
conventional s-wave superconductor with singlet pairing. The Hamiltonian
describing the system can be written as
\begin{equation}
H=H_{0}-\sum_{\{p,s\}}\{a_{ps}^{\dagger }J\hat{\sigma}_{3}a_{ps^{\prime
}}+(\Delta a_{\bar{p}\bar{s}}^{\dagger }a_{ps}^{\dagger }+\mathrm{c.c.})\}\;,
\label{e1}
\end{equation}
where $H_{0}$ is the one-particle electron energy including interaction with
impurities, $J$ is the exchange field which is nonzero only inside the F
grain, and $\Delta $ is the superconducting order parameter which vanishes
inside F. The matrix $\hat{\sigma}_{3}$ is the Pauli matrix in the
spin-space and we have assumed that the magnetization of the grain is in the
$z$ direction. The notation $\bar{p}$ and $\bar{s}$ means inversion of
momentum and spin respectively.

In order to calculate physical quantities of interest it is convenient to
use quasiclassical Green's functions $\check{g}$ and write the proper
quasiclassical equations for them. We consider the ``dirty'' limit, which
means that the inverse momentum relaxation time $\tau ^{-1}$ is much larger
than the exchange energy $J$ and than the critical temperature $T_{c}$ of
the superconductor. The functions $\check{g}$ are $4\times 4$ matrices in
the particle-hole and spin space and obey the Usadel equation which in the
presence of the exchange interaction has the form
\begin{equation}
D\nabla \left( \mbox{$\check{g}$}\nabla \check{g}\right) -\omega \left[ %
\mbox{$\hat{\tau}_3$}\mbox{$\hat{\sigma}_0$},\mbox{$\check{g}$}\right] +iJ%
\left[ \mbox{$\hat{\tau}_3$}\mbox{$\hat{\sigma}_3$},\mbox{$\check{g}$}\right]
=-i\left[ \check{\Delta},\mbox{$\check{g}$}\right] \;.  \label{Usad}
\end{equation}
In the $S$ region $D=D_{S}$, $J=0$, $\check{\Delta}=\Delta i%
\mbox{$\hat{\tau}_2$}\mbox{$\hat{\sigma}_3$}$ (the phase of $\Delta $ is
chosen to be zero). In the $F$ grain $D=D_{F}$ and $\Delta =0$. Eq. (\ref
{Usad}) should be complemented by boundary conditions at the $S/F$ interface
\cite{Zaitsev}
\begin{equation}
\gamma _{F}\left( \mbox{$\check{g}$}\mathbf{n}\nabla \check{g}\right)
_{F}=\gamma _{S}\left( \mbox{$\check{g}$}\mathbf{n}\nabla \check{g}\right)
_{S};\ \gamma _{F}\left( \mbox{$\check{g}$}\mathbf{n}\nabla \check{g}\right)
_{F}=-\left[ \mbox{$\check{g}$}_{S},\mbox{$\check{g}$}_{F}\right] ,\;\;\;
\label{bc2}
\end{equation}
where $\gamma _{S,F}=\sigma _{S,F}R_{b}$, $\sigma _{S,F}$ are the
conductivities of the $F$ and $S$ layers, $R_{b}$ is the resistance per unit
area, and $\mathbf{n}$ is the unit vector normal to the $S/F$ interface.
Solving Eqs. (\ref{Usad}, \ref{bc2}) for the Green functions $\check{g}$, one
can obtain the induced magnetization $\delta M$
\begin{equation}
\delta M=\mu \delta \sum_{p}(<c_{p\uparrow }^{\dagger }c_{p\uparrow
}-c_{p\downarrow }^{\dagger }c_{p\downarrow }>)=-\mu i\pi \nu T\sum_{\omega
=-\infty }^{\omega =+\infty }Tr(\hat{\sigma}_{3}\hat{g})/2  \label{M}
\end{equation}
where $\mu $ is an effective Bohr magneton, $\nu $ is the density of states
(DOS) and the sum is taken over the Matsubara frequencies $\omega =\pi
T(2n+1)$.

The functions $\check{g}$ can be represented in the form $\check{g}=i\hat{%
\tau}_{2}\hat{f}+\hat{\tau}_{3}\hat{g}$, where the condensate function $\hat{%
f}$ and the normal function $\hat{g}$ are matrices in the spin space \cite
{BVE1}. For simplicity we consider the case of a small grain when the
condition $a\leq \xi _{F}=\sqrt{D_{F}/J},$ fulfilled (this is possible
provided the exchange energy $J$ is much smaller than the Fermi energy $%
\varepsilon _{F}$). In this limit the solution can be found averaging Eq. (%
\ref{Usad}) over the grain volume and we obtain for the Green functions
\begin{equation}
g_{F\pm }=\tilde{\omega}_{\pm }/\zeta _{\omega \pm },\;\;f_{F\pm }=\pm
\epsilon _{bF}f_{BCS}/\zeta _{\omega \pm }  \label{gfPM}
\end{equation}
Here $g_{F\pm }$ and $f_{F\pm }$ are the diagonal elements of the matrices $%
\hat{g}$ and $\hat{f}$, $\tilde{\omega}_{\pm }\!\!=\!\!\omega +\epsilon
_{bF}g_{BCS}\mp iJ$, $\zeta _{\omega \pm }=\sqrt{\tilde{\omega}_{\pm
}^{2}-(\epsilon _{bF}f_{BCS})^{2}}$, $g_{BCS}=i(\omega /\Delta
)f_{BCS}=\omega /\sqrt{\omega ^{2}+\Delta ^{2}},\ \epsilon
_{bF}=3D_{F}/(2\gamma _{F}a)$. When obtaining Eq.(\ref{gfPM}), we assumed
that the functions $\hat{g}_{S},$ $\hat{f}_{S}$ are close to their BCS
values $\hat{g}_{BCS},$ $\hat{f}_{BCS}$, i.e. the corrections $\delta \hat{g}%
_{S},$ $\delta \hat{f}_{S}$ are small. Under this assumption Eq.(\ref{Usad})
for the functions $\delta \hat{g}_{S},$ $\delta f_{S}$ can be linearized.
For example, the linearized Usadel equation for $\delta g_{S3}\equiv Tr(\hat{%
\sigma}_{3}\hat{g})/2$ has the form
\begin{equation}
\nabla ^{2}\delta g_{S3}-\kappa _{S}^{2}\delta g_{S3}=0  \label{e2}
\end{equation}
The function $\delta g_{S3}$ determines the excess magnetization. Solving
Eq. (\ref{e2}) with the boundary condition, Eq. (\ref{bc2}), we obtain
\begin{equation}
\delta g_{S3}=\frac{f_{BCS}}{\gamma_S}\left(g_{BCS}f_{F0}-f_{BCS}g_{F3}\right)\frac{a^2}{1+\kappa_S a}\frac{e^{-\kappa_S(r-a)}}{r}\;,
\label{gs3}
\end{equation}
where $\kappa _{S}^{2}=2\sqrt{\omega ^{2}+\Delta ^{2}}/D_{S},$ $%
f_{F0}=(f_{F+}-f_{F-})/2$ describes the triplet component mentioned above.
One can see that the correction $\delta g_{S3}$ is small if the parameter $%
a/\gamma _{S}$ is small. Note also that the function $\delta g_{S3}$, which
according to Eq.(\ref{M}) determines the induced magnetization, is
proportional to the condensate function $f_{F0}$. The latter function is the
triplet component with zero projection of the magnetic moment on the $z$
axis. Using Eq.(\ref{gs3}) one can easily calculate the total magnetic
moment of the $S$ region
\begin{equation}
\mathcal{M}_{S}=-i\pi \nu_S T\mu \sum_{\omega =-\infty }^{\omega =+\infty
}\int d^{3}r\delta g_{S3}\; .
\label{totms}
\end{equation}
It is not difficult to see that the magnetic moment $M_{S}$ has the sign
opposite to $J$ (the magnetic moment of the ferromagnetic particle $M_{F0}$
is proportional to $J$), which means that the spins of the conduction
electrons screen (at least partially) the magnetic moment $M_{F0}$ of the
ferromagnetic grain.

Let us compare the total magnetic moment induced in the superconductor, Eq.(%
\ref{totms}), with the magnetic moment of the ferromagnetic grain in the
normal state $\mathcal{M}_{F0}=(4\pi a^{3}/3)M_{F0}.$ The sum in Eq.(\ref
{totms}) can be computed numerically in a general case. For simplicity we
consider here a limiting case that can be realized experimentally.

We assume that the transmission coefficient through the $S/F$ interface
 is not small and the condition  $\Delta <<J\leq (D_{F}/a^{2})$ is
fulfilled. In this case the expression for $f_{F0}$ is drastically
simplified. To estimate the energy $D_{F}/a^{2}$ we assume that the mean
free path is of the order of $a$. For $a=30\AA $ and $v_{F}=10^{8}cm/\sec $
we get $D_{F}/a^{2}\approx 1000K;$. This condition is fulfilled for
ferromagnets with the exchange energy of the order of several hundreds $K$.
In order to relate $M_{F0}$ to $J$, one has to make a certain assumption
about the nature of the ferromagnet. If the magnetic moment $M_{F0}$ is
induced mainly by free electrons (an itinerant ferromagnet), one gets $%
M_{F0}=\mu \nu _{F}J.$ Then we obtain for low temperatures
\begin{equation}
\mathcal{M}_{S}/\mathcal{M}_{F0}=-1\;.  \label{e3}
\end{equation}
Eq. (\ref{e3}) describes a remarkable phenomenon: at sufficiently low
temperatures and in the limit of a small grain ($a\leq \xi _{F}$) the
magnetic moment of the latter is screened over distances of the order of $%
\xi _{S}$. This screening is complete if the magnetization of the
$F$ particle is due to the free electrons (itinerant ferromagnet).
It can be easily shown that at arbitrary temperatures this ratio
is equal to $-(1-n_{n}(T)/n_{e})$, where $n_{e}$  and $n_{n}(T)$
are the density of total number of electrons and "normal"
electrons defined in Ref.\cite{abri_book}.

The compensation of the magnetization of an itinerant ferromagnet by the
Cooper pairs is to some extent consistent with the result obtained by
Rusinov and Gor'kov some decades ago\cite{gorkov&rusinov}. They studied the
properties of a superconductor with paramagnetic impurities that were
assumed to be ferromagnetically ordered. The free electrons interact with
the  magnetic impurities via the exchange interaction. Their
approach (averaging over impurity positions) reduces the problem to finding
the magnetization of a superconductor with an effective exchange interaction
uniformly distributed in space. It was demonstrated in Ref. \cite{gorkov&rusinov} that the total itinerant magnetization of the system was
zero in the limit of low temperatures and not too large exchange energy.

Clearly, the screening obtained here is due to the appearance of the triplet
component. So, if the latter is suppressed by other mechanisms the effect
will be reduced. The spin-orbit interaction (SOI) and orbital effects
(Meissner currents) are such mechanisms. The other way of thinking of this
reduction is that the spins of the electrons of the Cooper pairs are not
necessarily antiparallel in the presence of e.g. SOI and are not as
efficient in inducing the magnetization.

In Ref.\cite{BVE1} we studied the effect of the SOI on the triplet component
penetration into the ferromagnet. Here we can simply use these results to
analyze the effect of the SOI on the penetration of the induced
magnetization into the superconductor.

If the SOI is taken into account, an additional term of the form $(i/\tau
_{s.o.})\left[ \check{S}\mbox{$\hat{\tau}_3$}\mbox{$\check{g}$}%
\mbox{$\hat{\tau}_3$}\check{S},\mbox{$\check{g}$}\right] $ appears in the
Usadel equation, where $\tau _{s.o.}$ is the spin-orbit
scattering time, $S=(\sigma _{1},\sigma _{2},\sigma _{3}\tau _{3})$ (see
\cite{demler} and \cite{BVE1}). Due to this additional term the quantity $%
\kappa _{S}^{2}$ in the linearized Usadel equation is replaced by
\begin{equation}
\kappa _{S}^{2}=\kappa _{S}^{2}+\kappa _{so}^{2}\; ,  \label{e4}
\end{equation}
where $\kappa_{so}^{2}=8D_{S}/\tau _{so}$.  Therefore the length of the
penetration of $f_{S0}$ and of $M_{S}$ into the $S$ region decreases if $%
\kappa _{S}^{2}\sim \xi _{S}^{-2}<\kappa _{so}^{2}.$ In principle, one can
measure the spatial distribution of the magnetic moment in the $S$ region
(see e.g. \cite{Suter} ) and get an information about the SOI in the
superconductors. With the help of Eq. (\ref{e4}), this \ would be an
alternative method to measure the strength of the SOI in superconductors,
complementary to the measurement of the Knight shift \cite{Knight}. (The
latter is based on the result of Ref. \cite{Abr&Gor} that the Knight shift
observed in superconductors is due to the SOI). Let us notice, that in the
presence of  SOI the linearized equation for the function $f_{S3}$ that
determines a correction to the energy gap due to the proximity effect
remains unchanged.

The orbital effects (the Meissner currents) also change the characteristic
length of the penetration of the induced magnetization $M_{S}$. Contrary to
the case of bulk conventional superconductors, the Meissner currents in $S/F$
structures arise spontaneously even in the absence of an external magnetic
field $H_{ext}$. This happens because an internal magnetic field is induced by
the ferromagnet. These spontaneous currents were studied, e.g., in Refs.
\cite{BVE3,Annett}. In Ref. \cite{BVE3} the spatial dependence of the
Meissner current in the $F$ film was calculated, whereas the Meissner
currents in the both $S$ and $F$ films were computed in Ref. \cite{Annett}.
It is of interest to know not only the spatial dependence and the magnitude
of the Meissner currents, but also their effect on the penetration length of
the $TC$ and on the induced magnetization. Here we study the influence of
the Meissner currents on the penetration length of $M_{S}.$ For simplicity
we consider first a planar geometry, which allows us to get a simple
solution. For a spherical particle we estimate the effect by order of
magnitude.

\begin{figure}[tbp]
\onefigure[scale=0.5]{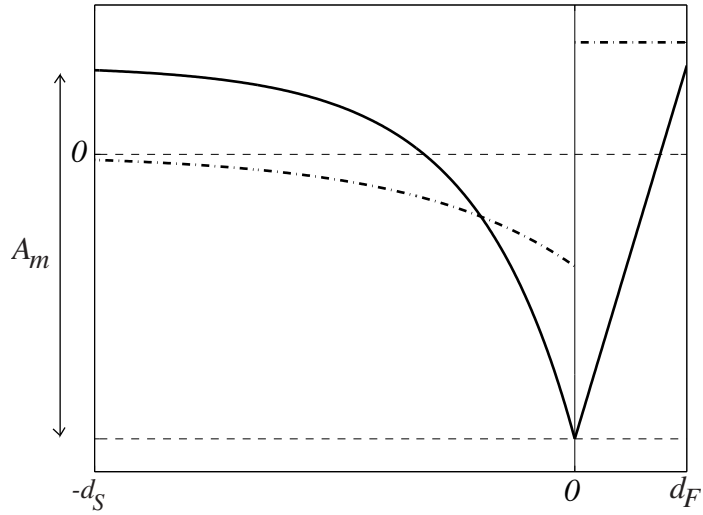}
\caption{Spatial dependence of the vector potential $A(x)$ or the supercurrent density. For low enough
temperatures the values of $A$ at both outer surfaces ($x=-d_s$ and $x=d_F$)
is nearly the same. The dot-dashed line shows the spatial dependence of the
local magnetization in arbitrary units}
\label{fig.2}
\end{figure}

Let us consider a bilayer $S/F$ structure with the thicknesses $d_{S,F}$. {%
We assume again that the thickness of the }${F}${\ layer is small (}$%
d_{F}<<\xi _{F}${\ ) and the thickness of the }${S}${\ layer obeys the
condition: }$\xi _{S}<<${\ \ }$d_{S}<<\lambda _{L}.${\ \ In this case the
solution for }$g_{F\pm }${\ \ and }$f_{F\pm }${\ \ (Eq.(\ref{Usad})) remains
unchanged provided one replaces }$a/3${\ \ with }$d_{F}${\ . The solution
for }$\delta g_{S3}${\ has the form }$\delta g_{S3}=(\kappa
_{S}g_{BCS}f_{BCS}/\gamma _{S})f_{F0}\exp (-\kappa _{S}x)$ and {one can
easily calculate the spatial distribution of the magnetization in the system
}$M_{S,F}(x)$. Note that again in the case of an itinerant ferromagnet and
low temperatures the total magnetic moment in S is equal to $(-$ $%
M_{F0}d_{F})$. We display schematically the spatial dependence of the
magnetization in Fig.2 alongside with the spatial dependence of the vector
potential $A(x)$ {\ which is given by}
\begin{equation}
A(x)=A_{0}+4\pi \int_{0}^{x}dx^{\prime }M(x^{\prime }).
\end{equation}
{with a constant }$A_{0}${\ determined from the condition that
the total current through the system is zero (no external magnetic field). We have assumed that the magnetization lies in-plane. The
supercurrent density is expressed through }$A$ as
\begin{equation}
j(x)=2\pi T\sigma A(x)/(e\phi _{0})\sum_{\omega }f_{3}^{2}
\end{equation}
where $\phi _{0}${\ \ is the magnetic flux quantum. In the limit
}$J<<\epsilon _{bF}$ {\ the condensate functions }$f_{3}^{2}${\ \ are nearly
the same in the }${S}${\ and }${F}${\ layers. Taking this into account and
calculating the total current }$I$, {\ we get from the condition }$I=0$ for $%
A_{0}$
\begin{equation}
A_{0}=-A_{m}\left( \sigma _{S}d_{S}+\sigma _{F}d_{F}/2-\sigma _{S}\frac{2\pi
T}{D_{s}}\sum_{\omega }\frac{f_{BCS}^{2}}{\kappa _{S}^{3}}\right) /(\sigma
_{S}d_{s}+\sigma _{F}d_{F})\;.
\end{equation}
where $A_{m}=4\pi M_{F0}d_{F}${\ . At low temperatures the third term is
approximately equal to }$0.53\sigma _{S}\sqrt{D_{S}/2\Delta }$. {\ The
spatial dependence of the vector potential is shown in Fig. \ref{fig.2}. Eq.(%
\ref{gs3}) for }$\delta g_{S3}${\ \ does not change in the presence of the
Meissner current provided one replaces }$\kappa _{S}^{2}${\ \ with }$\kappa
_{S}^{2}+p_{0}^{2}${, where }$p_{0}=A_{0}/\phi _{0}$.{\ \ It is clear that
the orbital effects are negligible if }$A_{0}/\phi _{0}\cong A_{m}/\phi
_{0}<<\xi _{S}^{-1}$. For example, $A_{m}/\phi _{0}\sim 5.10^{3}$cm$^{-1}$
for $4\pi M_{F}\sim 1$kOe and $d_{F}\sim 50$\AA , and therefore $\xi _{S}$
should be smaller than $2\mu $m.

It is not difficult to estimate $p_{0}$ in the case of a spherical particle.
The quantity $p_{0}$ in this case reaches the maximum at $r=a$ and is of the
order of $\ (4\pi M_{F}/\phi _{0})a.$
%%%%%%%%%%%%%%%%%%%%%%%%%%%%%%%%%%
%%%%%%%%%%%%%%%%%%%%%%%%%%%%%%%%%%%%
As we have noted above, the magnetic moment in $F$ is compensated by that in
the superconductor. This means in particular that the total internal
magnetic flux in the $S/F/S$ Josephson junction may be zero even if the $F$
layer is a single domain (see experiments of Ref. \cite{Ryaz}).

In summary, we have shown that the magnetic moment $M_{F0}$ of a
ferromagnetic particle embedded into a superconductor is screened by spins
of the Cooper pairs. An induced magnetic moment $M_{S}$ aligned in the
opposite to $M_{F0}$ direction arises in the superconductor.

The induced magnetic moment in the superconductor can be observed
experimentally. In a recent experiment \cite{Suter} the spatial electron
spin polarization was determined by means of muon spin rotation. Such a
method may be used in order to determine $M_{S}$.
% This can be done in a similar way as the spatial
%distribution of the magnetic field in a superconductor has been measured
%with the help of measurements of the muon spin precession \cite{Suter}.
Another possible method is the measurement of the Knight shift in
superconductors that should be dependent on the magnetization $M_{S}$.One can also determine the total magnetic moment of a  S/F structure performing magnetic resonant measurements as in Ref. \cite{Garif}.

At last, one can determine $M_{S}$ measuring the total magnetic moment of a
superconductor with embedded ferromagnetic particles similarly to the work
\cite{Gerber}, where an enhancement of the magnetic moment of ferromagnetic
particles embedded into a nonmagnetic matrix has been observed. Although the
enhancement in the case of a normal metal matrix awaits its explanation, we
suggest to measure the magnetization of the system of ferromagnetic
particles embedded in a superconducting matrix. When using a superconducting
metal as the matrix, a reduction of the magnetic moment, instead of the
enhancement, should be observed below the superconducting temperature $T_{c}$%
.

We would like to thank SFB 491 \textit{Magnetische Heterostrukturen} for
financial support.


\begin{thebibliography}{0}

%\bibitem{Gerber}  A.Gerber, A.Milner, M.Karpovshi, and A.Tsukernik, Evidence
%of the Temperature Dependent Conduction Electron Spin Polarization in
%Nanoscale Ferromagnet/Normal-Metal Systems, Preprint (2003).

\bibitem{abri_book}  \Name{ Abrikosov A. A.}
\Book{Fundamentals of the Theory of
Metals} \Publ{North-Holland, Amsterdam} \Year{1988}.

\bibitem{Abr&Gor}  \Name{Abrikosov A. A.\and Gorkov L.P.} \REVIEW{Sov. Phys.
JETP}{15}{1962}{752}.

\bibitem{BVE1}  \Name{Bergeret F. S., Volkov A. F. \and Efetov K. B.} %
\REVIEW{Phys. Rev. B}{68}{2003}{064513}.

\bibitem{Zaitsev}  \Name{ Zaitsev A. V.} \REVIEW{Sov. Phys. JETP}
{59}{1984}{863}; \Name{Kuprianov M. Y. \and  Lukichev V. F.} \REVIEW{ Sov.
Phys. JETP}{64}{1988}{139}.

\bibitem{gorkov&rusinov}  \Name{Gor'kov L. P. \and  Rusinov A. L.} %
\REVIEW{Sov. Phys. JETP} {19}{1964}{922}.

\bibitem{demler}  \Name{Demler E. A.,  Arnold G. B. \and Beasley M.R} %
\REVIEW{Phys. Rev. B}{55}{1997}{15174}.

\bibitem{Suter}  \Name{Luetkens H. et al.} \REVIEW{Phys. Rev.
Lett.}{91}{2003}{017204}; \Name{Suter A. et al.} %
\REVIEW{cond-mat/0310203}{}{2003}{}

\bibitem{Knight}  \Name{Androes G. M. \and Knight W. D.} \REVIEW{Phys.
Rev.}{121}{1961}{779}.

\bibitem{BVE3}  \Name{Bergeret F. S., Volkov A. F. \and Efetov K. B.} %
\REVIEW{Phys. Rev. B}{64}{2001}{134506}.

\bibitem{Annett}  \Name{Krawiec M. et al.} \REVIEW{Phys. Rev.
B}{66}{2002}{172505}.

\bibitem{Ryaz}  \Name{Ryazanov V. V. et al.} \REVIEW{Phys. Rev.
Lett.}{86}{2001}{2427}; \Name{Kontos T. et al.} \REVIEW{Phys. Rev.
Lett.}{89}{2002}{137007}; \Name{Blum Y. et al.} \REVIEW{Phys. Rev.
Lett.}{89}{2002}{187004}.


%A.Milner, M.Karpovshi, and A.Tsukernik,
%Evidence
%of the Temperature Dependent Conduction Electron Spin Polarization in
%Nanoscale Ferromagnet/Normal-Metal Systems, Preprint (2003).

%\bibitem{muhge}  Th. Muehge et al.; Phyisica C {  296}, 325 (1998).

\bibitem{Garif}
 \Name{Garifullin I.A.  et al.}
 \REVIEW{Appl.Magn.Reson.}{22}{2002}{439}.

\bibitem{Gerber}  \Name{Gerber A. et al}
\REVIEW{Evidence
of the Temperature Dependent Conduction Electron Spin Polarization in
Nanoscale Ferromagnet/Normal-Metal Systems, Preprint}{}{2003}{}.
%\bibitem{Suter}  A. Suter et al., cond-mat/0310203.

%\bibitem{AG_Knight}  A. A. Abrikosov, L. P. Gorkov, Sov.Phys. JETP {\bf 15},
%752 (1962).
%\end{references}
\end{thebibliography}
\end{document}